\documentclass[a4paper,11pt]{article}
\pdfoutput=1 % if your are submitting a pdflatex (i.e. if you have
\usepackage{jheppub,bm,mathtools,slashed} 
\usepackage[usenames,dvipsnames,svgnames,table,x11names]{xcolor}
\usepackage{extarrows}
\usepackage{easybmat}

\usepackage{nicematrix}
\usepackage[whole]{bxcjkjatype} 
\usepackage[T1]{fontenc} % if needed
\usepackage[utf8]{inputenc}
\usepackage{amsmath,amssymb}
\usepackage{amsthm}
\usepackage{bm}
\usepackage{graphicx}
\usepackage{ascmac}
\usepackage{enumerate}
\usepackage{tensor}
\usepackage{physics}
\usepackage{color}
\usepackage{slashed}
\usepackage{braket}
\usepackage{url}
\usepackage{mathrsfs}
\usepackage{comment}

\renewcommand*{\backref}[1]{}
\renewcommand*{\backrefalt}[4]{%
  \ifcase #1 %
    \relax % use \relax if you do not want the "No citations" message
  \or
    {\scriptsize (page~#2).}%
  \else
    {\scriptsize (pages~#2).}%
  \fi%
}

\definecolor{light_blue}{rgb}{0.15, 0.35, 0.95}
\definecolor{kit_green}{rgb}{0, 
0.58823 %150/255
, 0.50980 %130/255
}

\usepackage{tikz}
\usepackage{pgfplots}
\pgfplotsset{compat=1.14}
\usepackage{tikz-3dplot}
\usetikzlibrary{intersections, calc,positioning,decorations.pathreplacing,decorations.pathmorphing,arrows,arrows.meta,patterns,pgfplots.fillbetween,math,matrix}

\usepackage{multirow}
\usepackage{booktabs} %for \toprule
\usepackage{siunitx} %for \num
\usepackage{tabularx} %for tabularx

\excludecomment{note}

\title{
A Casimir obstruction to asymptotically flat
black-brane completions of non-supersymmetric 7-branes
}

\author[1,2,3]{Yuta Hamada,}
\author[2]{Hayate Kimura,}

\affiliation[1]{Theory Center, IPNS, High Energy Accelerator Research Organization (KEK), 1-1 Oho, Tsukuba, Ibaraki 305-0801, Japan}
\affiliation[2]{Graduate Institute for Advanced Studies, SOKENDAI, 1-1 Oho, Tsukuba, Ibaraki 305-0801, Japan}
\affiliation[3]{RIKEN Center for Interdisciplinary Theoretical and Mathematical Sciences (iTHEMS), RIKEN, Wako 351-0198, Japan}

\abstract{
We study axisymmetric 7-brane solutions in the dilaton-gravity sector of 10d supergravity, including the backreaction of the Casimir energy induced by monodromies around the transverse circle. Without Casimir energy, we find that the system is analytically solvable and admits locally flat asymptotics with arbitrary deficit angles, but the corresponding solutions contain a naked singularity at finite proper distance. We investigate whether this singularity can be replaced by a regular finite-horizon black-brane core once Casimir backreaction is included.
We find that, within our ansatz, Casimir backreaction obstructs an asymptotically locally flat black-brane completion of the naked 7-brane solution.
}

\preprint{KEK-TH-2841,~RIKEN-iTHEMS-Report-26}

\begin{document}
\maketitle

\section{Introduction}\label{sec:intro}

Codimension-two branes provide an important arena for studying symmetry structures of string theory. 
The backreaction is not localized in a rapidly decaying way, and the geometry at infinity can retain direct information about the charge and monodromy of the brane. 
A familiar example is the D7-brane in type IIB string theory, around which the axio-dilaton undergoes an $SL(2,\mathbb{Z})$ monodromy \cite{Greene:1989ya}. 
More generally, string theory admits branes associated with discrete duality transformations, including exotic branes~\cite{deBoer:2012ma}, and such objects are expected to appear as boundaries of lower-dimensional duality-twisted backgrounds. 
The existence of such codimension-two branes is closely related to the question of whether the monodromies of exotic branes can be observed in flat spacetime, as discussed recently in \cite{Sen:2025iuf}.

Recent developments have emphasized this viewpoint from the perspective of the cobordism conjecture~\cite{McNamara:2019rup}. 
If a compactification with a nontrivial holonomy represents a nontrivial cobordism class, the conjecture suggests that there should exist a corresponding defect or brane ending the background. 
This reasoning predicts, for example, reflection 7-branes in type II string theory \cite{Dierigl:2022reg,Debray:2023yrs} and exotic non-supersymmetric heterotic 7-branes associated with disconnected components of the gauge group \cite{Kaidi:2023tqo,Kaidi:2024cbx}. 
Understanding the spacetime geometry sourced by such branes is therefore important for clarifying how these cobordism charges are realized in supergravity.
For heterotic branes of lower spatial dimensions, gravitational solutions have already been constructed: the heterotic 6-brane solution was obtained in \cite{Horowitz:1991cd}, heterotic 0 and 4-brane solutions were constructed more recently in \cite{Fukuda:2024pvu}, and analytic solutions with a non-trivial temporal component of the gauge field were obtained in \cite{Chikazawa:2025cis}.

This work builds on our previous analysis of gravitational solutions in the presence of Casimir energies in symmetry-twisted backgrounds~\cite{Hamada:2025duq}. 
In the presence of a nontrivial monodromy around a 7-brane, the fields acquire boundary conditions along the transverse circle, and the resulting Casimir energy can contribute to the effective stress tensor. 

In this paper we study axisymmetric supergravity solutions describing such 7-branes. 
We restrict attention to the dilaton-gravity sector and set all gauge-field configurations to constants, which are zero in most of the examples considered below. 
Within this ansatz, the 10d problem reduces to 3d Einstein gravity coupled to scalar fields. 
A key distinction is whether the 9d background obtained by compactifying around the brane preserves supersymmetry. 
When it does, the Casimir energy around the transverse circle cancels. 
When it does not, the Casimir energy is nonzero and its gravitational backreaction must be included.

We first analyze the equations in the absence of Casimir energy. 
In this case the system is analytically solvable, and we find a family of solutions whose asymptotic geometry can be locally flat with an arbitrary deficit angle.
The solutions have a naked singularity at finite proper distance. 
These solutions describe the far-region behavior of 7-branes whose associated circle compactification preserves supersymmetry, such as the type II $(-1)^{F_L}$ reflection 7-brane and the heterotic 7-brane with periodic spin structure.

We then include the Casimir energy, corresponding to 7-branes associated with non-supersymmetric circle compactifications, such as the heterotic 7-brane with antiperiodic spin structure.
In this setup, we ask whether the naked singularity can be replaced by a regular finite-horizon black-brane core.
Although the resulting equations are no longer analytically solvable, they possess scaling properties that allow us to reduce the space of regular black-brane solutions to a universal canonical solution. 
By constructing this solution numerically and comparing its far-region behavior with the analytic solutions without Casimir energy, we find that the asymptotic geometry is driven to a non-flat branch. 
In particular, within our ansatz, nonzero Casimir energy obstructs the existence of asymptotically locally flat black-brane completions of the 7-brane solutions.

The rest of this paper is organized as follows. 
In Sec.~\ref{sec:eom}, we introduce the axisymmetric ansatz and derive the equations of motion with and without Casimir energy. 
In Sec.~\ref{sec:solutions}, we solve the equations analytically in the case without Casimir energy, and construct the universal Casimir-supported solution numerically. 
In Sec.~\ref{sec:sols}, we apply the general results to type II reflection 7-branes and heterotic 7-branes. 
We conclude with comments on possible extensions and loopholes.

%%%%%%%%%%%%%%%%%%%%%%%%%%%%%%%%%%%%%%%%%%%%%%%%%%%%%%%%%%%%%%%%%%%%%%%%%%%%%%
\section{Axisymmetric ansatz and equations of motion}\label{sec:eom}
In this section, we formulate the equations of motion governing axisymmetric 7-brane backgrounds. We first derive the equations of motion in the classical supergravity approximation. Dimensional reduction along the brane directions shows that the problem is equivalent to 3d Einstein gravity coupled to two massless scalar fields. 
We then incorporate the Casimir energy, which gives a quantum correction to the effective action and modifies the radial equations of motion. 
\subsection{Without Casimir energy}

We focus on 7-brane background solutions in which no gauge field in the theory acquires a nontrivial configuration. Therefore, our starting point is the dilaton-gravity sector of the supergravity action. 
\begin{align}
    S = \int d^{10} x \sqrt{-g}e^{-2\phi}[\mathcal{R}+4\partial_\mu\phi\partial^\mu\phi] .\label{eq:10dim_sugra_action}
\end{align}
This action is independent of the type of string theory. We consider the following string-frame metric ansatz, inspired by \cite{Horowitz:1991cd}:
\begin{align}
    ds^2 = e^{2\omega} \tilde{g}_{\mu\nu}dx^\mu dx^\nu + e^{2\psi}dx_i dx^i, \hspace{10pt} \mu, \nu = 0,1,2, \hspace{10pt} i = 3, \ldots, 9,
    \label{eq:metric_ansatz_string}
\end{align}
where $x^i$ and $x^\mu$ denote the coordinates along and transverse to the brane, respectively. The functions $\omega$ and $\psi$, as well as the components of the metric $\tilde{g}_{\mu\nu}$, depend only on $x^\mu$.
Under this ansatz, \eqref{eq:10dim_sugra_action} becomes (see Appendix \ref{sec:Appendix_1} for details)
\begin{align}
     S = \int d^{10}x \sqrt{-\tilde{g}}\,e^{-2\phi +\omega + 7\psi} \bigg[&\mathcal{\Tilde{R}}-4\Tilde{\nabla}^2\omega %\nonumber \\
    %& \hspace{30pt}
    -2\Tilde{\partial}_\mu\omega\Tilde{\partial}^\mu\omega 
      - 56\tilde{\partial}_\mu\psi\tilde{\partial}^\mu\psi  \nonumber \\
      &% \hspace{30pt} 
      - 14 \tilde{\nabla}^2 \psi - 14\Tilde{\partial}_\mu \psi \Tilde{\partial}^\mu \omega + 4\tilde{\partial}_\mu\phi\tilde{\partial}^\mu\phi\bigg]. \label{eq:10dim_sugra_action_with_ansatz}
\end{align}
Here, $\tilde{\mathcal{R}}$ denotes the Ricci scalar constructed from $\tilde{g}_{\mu\nu}$, and $\tilde{\partial}^\mu:=\tilde{g}^{\mu\nu}\tilde{\partial}_\nu$. Introducing the fields $\phi$ and $\sigma$ by
\begin{align}
    &\omega = \dfrac{\phi}{4} + \dfrac{\sqrt{7}}{4}\sigma, %\\
    &&\psi = \dfrac{\phi}{4}-\dfrac{\sigma}{4\sqrt{7}},
\end{align}
the action \eqref{eq:10dim_sugra_action_with_ansatz} becomes
\begin{align}
    S = \left(\int d^7x\right)\int d^{3} x \sqrt{-\tilde{g}}\left[\tilde{\mathcal{R}}-\dfrac{1}{2}(\tilde{\nabla} \phi)^2 - \dfrac{1}{2}(\tilde{\nabla} \sigma)^2  \right].
\end{align}
In this way, our problem reduces to solving the 3d Einstein equations with two massless scalar fields, $\phi$ and $\sigma$. % in the spacetime. 
Now we impose the following 3d metric ansatz.
\begin{align}
   d\tilde{s}^2  = - A^2(z) dt^2 + dz^2 + R^2(z)d\varphi^2,
\end{align}
where $z$ is the proper distance from the brane, and $\phi$ and $\sigma$ are also functions of $z$. In this ansatz, the Einstein equations are given explicitly as follows:
\begin{align}
    &A^2 \left(\dfrac{1}{4}\sigma^{\prime 2} + \dfrac{1}{4}\phi^{\prime 2} + \dfrac{R^{\prime\prime}}{R}\right) = 0, \nonumber\\
    &\dfrac{A^\prime R^\prime}{AR} - \dfrac{1}{4}\sigma^{\prime 2} - \dfrac{1}{4}\phi^{\prime 2} = 0, \nonumber\\
    &R^2\left(\dfrac{1}{4}\sigma^{\prime 2} + \dfrac{1}{4}\phi^{\prime 2}+ \dfrac{A^{\prime \prime}}{A}\right) = 0, \nonumber\\
    &\left(\dfrac{A^\prime}{A} + \dfrac{R^\prime}{R}\right)\phi^\prime + \phi^{\prime\prime} = 0, \nonumber\\
    &\left(\dfrac{A^\prime}{A} + \dfrac{R^\prime}{R}\right)\sigma^\prime + \sigma^{\prime\prime} = 0.
    \label{eq:eom_without_Casimir}
\end{align}
Here, the prime denotes the derivative with respect to $z$.

\subsection{With Casimir energy}

In general, the presence of the brane can modify the equations of motion at the quantum level.
At low energies, the modification is captured by the Casimir energy \cite{Arkani-Hamed:2007ryu,Hamada:2025duq}, which we discuss in this section. In closed string theory, this term corresponds to the genus-one contribution in the worldsheet genus expansion and therefore does not couple to the dilaton in the string frame. Consequently, the effective action with Casimir energy is
\begin{align}
    S = \int d^{10}x\sqrt{-g}\bigg[e^{-2\phi}\mathcal{R}+4e^{-2\phi}\partial_\mu \phi \partial^\mu \phi - V_{\text{Casimir}}\bigg]. \label{eq:10dim_sugra_action_with_Casimir}
\end{align}
The Casimir energy in a spacetime containing a 7-brane is expressed as a function only of the radial coordinate measured from the brane. Although the explicit construction is presented in Appendix \ref{sec:Appendix_2}, its argument is determined by the proper radius of the transverse circle in the Einstein frame.
We employ the same string-frame metric ansatz as in the previous subsection~\eqref{eq:metric_ansatz_string}: 
\begin{align}
    ds^2 = e^{\frac{\phi}{2}+\frac{\sqrt{7}}{2}\sigma}[-A^2(z)dt^2 + dz^2 + R^2(z)d\varphi^2] + e^{\frac{\phi}{2}-\frac{\sigma}{2\sqrt{7}}}dx^idx_i. \label{eq:metric_ansatz_string_frame}
\end{align}
In the Einstein frame, the metric is rescaled as 
\begin{align}
    ds^{(E)2} = e^{-\frac{1}{2}\phi}ds^2 = e^{\frac{\sqrt{7}}{2}\sigma}[-A^2(z)dt^2 + dz^2 + R^2(z)d\varphi^2] + e^{-\frac{\sigma}{2\sqrt{7}}}dx^idx_i.
\end{align}
Therefore, the Casimir energy is a function of $\sqrt{g_{\varphi\varphi}^{(E)}} = e^{\frac{\sqrt{7}}{4}\sigma} R$ and depends on the metric components $\sigma$ and $R$. Under the metric ansatz, as in the previous subsection, the action \eqref{eq:10dim_sugra_action_with_Casimir} becomes
\begin{align}
    S = \int d^{3} x \sqrt{-\tilde{g}}\left[\tilde{\mathcal{R}}-\dfrac{1}{2}(\tilde{\nabla} \phi)^2 - \dfrac{1}{2}(\tilde{\nabla} \sigma)^2   - e^{\frac{5}{2}\phi + \frac{\sqrt{7}}{2}\sigma}V_\text{Casimir}(e^{\frac{\sqrt{7}}{4}\sigma} R) \right]. \label{eq:10dim_sugra_action_with_Casimir_with_ansatz}
\end{align}
after integrating out the directions parallel to the brane. The equations of motion receive additional contributions from the energy-momentum tensor associated with the Casimir energy. 

The Casimir energy for massless fields is given by (see Appendix \ref{sec:Appendix_2})
\begin{align}
    &V_{\text{Casimir}}(e^{\frac{\sqrt{7}}{4}\sigma} R) 
    =\sum_{p} (-1)^{2s_p+1} \dfrac{768 n_p}{(2\pi)^{15}}\dfrac{e^{-\frac{5\sqrt{7}}{2}\sigma}}{R^{10}} \sum_{n=1}^\infty \dfrac{\cos(2\pi n \theta_p)}{n^{10}}, \label{eq:Casimir_energy_for_sugra}
\end{align}
where the sum over $p$ runs over the field content, $n_p$ denotes the number of real degrees of freedom, $s_p$ is the spin of the field, and $\theta_p = 0\,(1/2)$ corresponds to the (anti-)periodic boundary condition.

Applying the formula to the supergravity field contents, we find
\begin{align}
    &V_\text{Casimir}(e^{\frac{\sqrt{7}}{4}\sigma} R) = - \alpha \dfrac{e^{-\frac{5\sqrt{7}}{2}\sigma}}{R^{10}},
    &&\alpha:= \dfrac{768}{(2\pi)^{15}}\sum_{p} (-1)^{2s_p} n_p \sum_{n=1}^\infty \dfrac{\cos(2\pi n \theta_p)}{n^{10}},
    \label{eq:universal_form_of_Casimir}
\end{align}
where $\alpha$ depends on the type of supergravity and the flux-charge of the brane. 
We treat the Casimir energy in a local approximation. 
At each value of $z$, it is taken to be the flat-space one-loop vacuum energy on a circle with Einstein-frame proper radius $L(z)=e^{\sqrt{7}\sigma(z)/4}R(z)$. In doing so, we keep only the leading massless-field contribution proportional to $L^{-10}$ and neglect derivative, curvature, higher-derivative, and massive-string-state corrections.
This approximation is expected to be valid when the circle is large compared with the string scale and the background varies slowly.

The equations of motion derived from the action \eqref{eq:10dim_sugra_action_with_Casimir_with_ansatz} with \eqref{eq:universal_form_of_Casimir} are
\begin{align}
    &A^2 \left(\dfrac{\alpha}{2} \dfrac{e^{-2\sqrt{7}\sigma + 5\phi/2}}{R^{10}} - \dfrac{1}{4}\sigma^{\prime 2} - \dfrac{1}{4}\phi^{\prime 2} - \dfrac{R^{\prime\prime}}{R}\right) = 0, \nonumber\\
    &-\dfrac{\alpha}{2} \dfrac{e^{-2\sqrt{7}\sigma + 5\phi/2}}{R^{10}} + \dfrac{A^\prime R^\prime}{AR} - \dfrac{1}{4}\sigma^{\prime 2} - \dfrac{1}{4}\phi^{\prime 2} = 0, \nonumber\\
    &R^2\left(\dfrac{9\alpha}{2} \dfrac{e^{-2\sqrt{7}\sigma + 5\phi/2}}{R^{10}} + \dfrac{1}{4}\sigma^{\prime 2} + \dfrac{1}{4}\phi^{\prime 2}+ \dfrac{A^{\prime \prime}}{A}\right) = 0, \nonumber\\
    &\dfrac{5\alpha}{2} \dfrac{e^{-2\sqrt{7}\sigma + 5\phi/2}}{R^{10}} + \left(\dfrac{A^\prime}{A} + \dfrac{R^\prime}{R}\right)\phi^\prime + \phi^{\prime\prime} = 0, \nonumber\\
    &-2\sqrt{7} \alpha \dfrac{e^{-2\sqrt{7}\sigma + 5\phi/2}}{R^{10}} + \left(\dfrac{A^\prime}{A} + \dfrac{R^\prime}{R}\right)\sigma^\prime + \sigma^{\prime\prime} = 0.
    \label{eq:eom_with_Casimir}
\end{align}
For $\alpha=0$, these reduce to \eqref{eq:eom_without_Casimir}, as they should.

%%%%%%%%%%%%%%%%%%%%%%%%%%%%%%%%%%%%%%%%%%%%%%%%%%%%%%%%%%%%%%%%%%%%%%%%%%%%%%
\section{Solutions} \label{sec:solutions}

In this section, we solve the equations of motion derived in the previous section. 
We first consider the case without Casimir energy, where the system can be solved analytically. 
We then turn to the case with nonzero Casimir energy and ask whether the naked singularity in the analytic solutions can be replaced by a regular finite-horizon black-brane core.
Although the full equations are no longer analytically tractable, their scaling properties allow us to reduce the problem to a canonical solution, which we construct numerically. 
By comparing its far-region behavior with the analytic solutions without Casimir energy, we show that the Casimir backreaction obstructs asymptotically locally flat black-brane completions within our ansatz.

\subsection{Solutions without Casimir energy}
As we will see later, the Casimir energy vanishes for 7-branes that can be viewed as boundaries of 9d supersymmetric string vacua.
In this case, we need to solve the equations without the Casimir contribution \eqref{eq:eom_without_Casimir}.  
We find that the general solution to this system of equations can be obtained analytically as follows (see Appendix~\ref{sec:Appendix_3} for the derivation):\footnote{Compared with the expressions in Appendix~\ref{sec:Appendix_3}, we explicitly restore the integration constant corresponding to shifts of the $z$ coordinate.}
\begin{align}
    g_{\mu\nu} &= e^{\frac{\phi}{2}+\frac{\sqrt{7}}{2}\sigma}[-a_0^2 (z-z_0)^{2\gamma}dt^2 + dz^2 + r_0^2 (z-z_0)^{2(1-\gamma)}d\varphi^2] + e^{\frac{\phi}{2} - \frac{\sigma}{2\sqrt{7}}}dx_i dx^i \nonumber\\
    \phi &= \phi_0 + 2\sqrt{\gamma(1-\gamma)}\sin\delta\ln (z-z_0) \nonumber\\
    \sigma &= \sigma_0 + 2\sqrt{\gamma(1-\gamma)}\cos\delta\ln (z-z_0).
    \label{eq:sol_without_Casimir}
\end{align}
where $a_0, \,r_0 >0, \,z_0 \in \mathbb{R}, ~0 \leq \gamma \leq 1, ~0 \leq \delta < 2\pi$, and $\phi_0,\sigma_0 \in \mathbb{R}$ are integration constants. In particular, the $\gamma = 0$ solutions have locally flat 3d metrics.
This is one of our main results.

For $0<\gamma<1$, this solution has a naked singularity at $z=z_0$. Moreover, the dilaton runs to infinity or minus infinity as $z \to z_0$, depending on the sign of $\sin\delta$. 
In both cases, the effective field theory description breaks down near the singularity.
For $\gamma=0$, the scalar fields are constant, and the solution describes a locally flat geometry with a conical singularity at $z=z_0$.
Therefore, we view these solutions as reliable only in the region away from the singularity.

\subsection{Casimir-supported black-brane solutions}

On the other hand, when the 7-brane background is a boundary of a 9d non-supersymmetric string vacuum, the Casimir energy is nonzero and the equations of motion are given by \eqref{eq:eom_with_Casimir}.
Our goal in this subsection is to explore black-brane solutions of these equations.
The equations possess the following properties:
\begin{enumerate}[i)]
    \item Invariance under shifts of the coordinate $z$: $z \to z + z_0$ with $z_0 \in \mathbb{R}$.
    \item Invariance under rescalings of $A$: $A \to a_0A$ with $a_0 \in \mathbb{R}_{>0}$. 
    \item Invariance under simultaneous shifts of $\phi$ and $\sigma$: $\phi \to \phi + \phi_0, ~\sigma \to \sigma + \sigma_0, ~\phi_0,\sigma_0 \in \mathbb{R}$ and $\phi_0/\sigma_0=4\sqrt{7}/5$. 
    \item Scaling property in the following sense. 
    Suppose that 
    \begin{equation}
        A_{\alpha_0}(z),\quad R_{\alpha_0}(z),\quad \phi_{\alpha_0}(z),\quad \sigma_{\alpha_0}(z)
    \end{equation}
    is a solution with a nonzero value $\alpha=\alpha_0$. Then, 
    we can construct a family of solutions with arbitrary nonzero $\alpha$, together with a rescaling of $R$ and shifts of $\phi$ and $\sigma$, as follows:
    \begin{align}
        &A_{\alpha}(z)=a_0A_{\alpha_0}(cz),
        &&R_{\alpha}(z)=r_0R_{\alpha_0}(cz),\nonumber\\
        &\phi_{\alpha}(z)=\phi_{\alpha_0}(cz)+ \phi_0,
        &&\sigma_{\alpha}(z)=\sigma_{\alpha_0}(cz)+ \sigma_0,
        \label{eq:rescaled_solutions_with_Casimir}
    \end{align}
    where
    \begin{align}
        c=\frac{e^{-\sqrt{7}\phi_0 + \frac{5}{4}\sigma_0}}{r_0^5} \sqrt{\frac{\alpha}{\alpha_0}}.
        \label{eq:c_general}
    \end{align}
    This property implies that the 7-brane backgrounds supported by the Casimir energy within our ansatz take a universal form. 
\end{enumerate}

To solve the equations, we must impose appropriate boundary conditions. 
We first note that the system has eight integration constants.
In the following, we argue that, using the properties above, all regular finite-horizon solutions can be written in the form of \eqref{eq:rescaled_solutions_with_Casimir} in terms of a single solution satisfying appropriate boundary conditions, which we call the canonical solution. 
First, using property i), we can place the horizon at $z = 0$,\footnote{Strictly speaking, the horizon position may lie at infinity. This would be a potential loophole in our analysis. See also the discussion at the end of this section.
}
which implies
\begin{align}
    A(0) = 0. \label{eq:boundary_condition_1}
\end{align}
Assuming that the 3d metric takes the form $-f(r)^2 dt^2 + dr^2/f(r)^2 + R(r)^2 d\phi^2$ and that the proper distance $z$ is defined by $dr/f(r) = dz$, we obtain
\begin{align}
    R'(0) = 0. \label{eq:boundary_condition_2}
\end{align}
unless $dR/dr$ diverges at the horizon.

Next, we impose the regularity conditions for 10d curvature corresponding to the metric \eqref{eq:metric_ansatz_string_frame}. Since the Ricci scalar is given by
\begin{align}
    \mathcal{R}^{(10)} = &e^{-\frac{1}{2}\phi-\frac{\sqrt{7}}{2}\sigma}\left[-\left(\dfrac{2A^\prime R^\prime}{A R} + \dfrac{2A^{\prime \prime}}{A} + \dfrac{2R^{\prime\prime}}{R} \right) - \dfrac{1}{2}\sigma^{\prime 2}  - \dfrac{1}{2}\phi^{\prime 2} \right. \nonumber \\ 
    & \left. - \left(\dfrac{A^\prime (\sqrt{7}\sigma^\prime + 9\phi^\prime)}{2A} - \dfrac{R^\prime (\sqrt{7}\sigma^\prime + 9\phi^\prime)}{2R}  -\dfrac{\sqrt{7}}{2} \sigma^{\prime \prime} - \dfrac{9}{2}\phi^{\prime \prime}\right) - 4\phi^{\prime 2} \right].
\end{align}
The finiteness of the Ricci scalar at the horizon requires the following conditions:
\begin{align}
    A^{\prime \prime} \to 0 , \hspace{10pt} \sqrt{7}\sigma^\prime + 9\phi^\prime \to 0, \label{eq:10dim_Ricci_scalar}
\end{align}
where $\to0$ means that the quantity goes to zero at least as fast as $A$. Even after imposing these conditions~\eqref{eq:10dim_Ricci_scalar}, $\mathcal{R}^{(10)}_{\mu\nu}\mathcal{R}^{(10)\mu\nu}$ contains a possible divergent contribution:
\begin{align}
\mathcal{R}^{(10)}_{\mu\nu} \mathcal{R}^{(10) \mu\nu} = 
e^{-\sqrt7\,\sigma - \phi}
\Bigg[
& A'^2 \Bigg( -4 \dfrac{1}{R}\dfrac{R'}{A} \dfrac{\phi'}{A}
+
\dfrac{72}{7}\dfrac{\phi'^2}{A^2} \Bigg)
\nonumber\\
&+
2A' \dfrac{\phi'}{A}
\Bigg( -2\dfrac{A''}{A} -4\dfrac{R''}{R} 
-8\phi'^2 + \dfrac{3}{\sqrt{7}}\sigma'' + \phi''
\Bigg)
\Bigg] \nonumber \\
&+ (\text{regular part}).
\end{align}
Combining this with the conditions \eqref{eq:10dim_Ricci_scalar}, the regularity conditions imply

\begin{align}
    \phi'(0) = 0, \hspace{10pt} \sigma'(0) = 0. \label{eq:boundary_condition_3}
\end{align}
At this stage, four of the eight integration constants are fixed by the boundary conditions (\eqref{eq:boundary_condition_1},\eqref{eq:boundary_condition_2},\eqref{eq:boundary_condition_3}). 
Then, using property ii), we can set $A^\prime(0)=1$; arbitrary constant rescalings of this solution are again solutions.
Property iii) gives the flat direction associated with shifts of $\phi(0)$ and $\sigma(0)$, which can be taken into account simply by shifting the solution.
These observations indicate that we can impose the following boundary conditions without loss of generality:
\begin{align}
    &A'(0) = 1, \hspace{10pt} 2\sqrt{7}\phi(0) + \dfrac{5}{2}\sigma(0) = 0. \label{eq:canonical_boundary_condition_1}
\end{align}
The remaining two integration constants are parametrized by
\begin{align}
    R(0) \quad \text{and} \quad x_0:=-2\sqrt{7} \sigma(0) + \dfrac{5}{2}\phi(0), 
    \label{eq:remaining_integration_constants}
\end{align}
where $x_0$ is the combination orthogonal to the flat direction in property iii).
Finally, property iv) indicates that solutions with arbitrary values of $R(0)$ and $x_0$ can be written in the form of \eqref{eq:rescaled_solutions_with_Casimir}.
In other words, the possible black-brane solutions of our system are exhausted by the class of solutions obtained from a single solution through the rescalings in iv). Therefore, it is sufficient to construct a solution satisfying the following canonical boundary conditions in addition to \eqref{eq:boundary_condition_1}, \eqref{eq:boundary_condition_2}, \eqref{eq:boundary_condition_3}, and \eqref{eq:canonical_boundary_condition_1}.
\begin{align}
    &R(0) = 1, \hspace{10pt} x_0 = 0. \label{eq:canonical_boundary_condition_2}
\end{align}
We call the solution satisfying \eqref{eq:boundary_condition_1}, \eqref{eq:boundary_condition_2}, \eqref{eq:boundary_condition_3}, \eqref{eq:canonical_boundary_condition_1}, and \eqref{eq:canonical_boundary_condition_2} the canonical solution.
In the following, we denote the canonical solution by hatted quantities:
\begin{align}
    \left(\hat{A}_\alpha(z),\,\hat{R}_\alpha(z),\,\hat{\phi}_\alpha(z),\,\hat{\sigma}_\alpha(z)\right).
    \label{eq:canonical_solution}
\end{align}
Then, the general solution is
\begin{align}
    &A_{\alpha}(z)=a_0\hat{A}_{\alpha_0}(cz),
    &&R_{\alpha}(z)=r_0\hat{R}_{\alpha_0}(cz),\nonumber\\
    &\phi_{\alpha}(z)=\hat{\phi}_{\alpha_0}(cz)+ \phi_0,
    &&\sigma_{\alpha}(z)=\hat{\sigma}_{\alpha_0}(cz)+ \sigma_0,
    \label{eq:general_solution_Casimir}
\end{align}
where $c$ is given by \eqref{eq:c_general}.

We have numerically constructed such a black-brane solution satisfying the canonical boundary conditions at its horizon, as shown in Fig.~\ref{fig:canonical_solutions}. 
For the numerical solution shown below, we choose
\begin{align}
    \alpha
    =\dfrac{3}{\pi^{15}} \left(1 - \dfrac{1}{2^{10}}\right)\zeta(10),
\end{align}
where $\zeta(s)$ is the zeta function.
As we will see in the next section, this value of $\alpha$ is relevant for the heterotic brane.

\begin{figure}
    \centering
    \begin{minipage}{.48\linewidth}
        \centering
        \includegraphics[width=\linewidth]{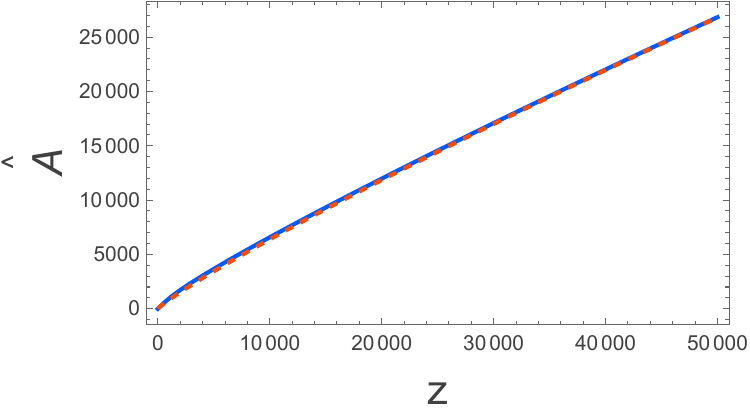}
    \end{minipage}
    \hfill
    \begin{minipage}{.48\linewidth}
        \centering
        \includegraphics[width=\linewidth]{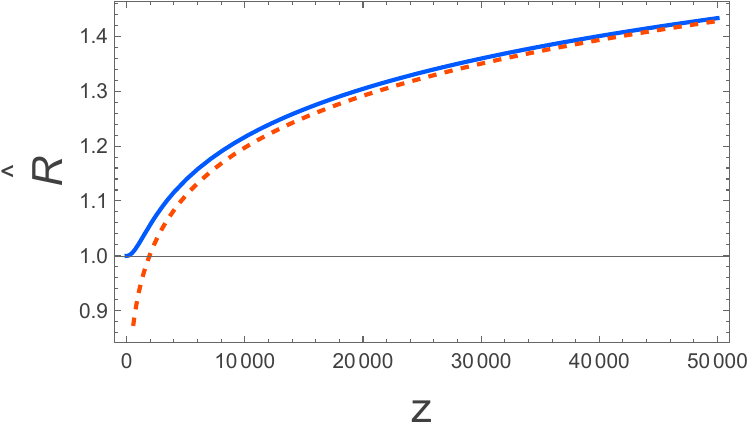}
    \end{minipage}
    \vspace{1em}
    \begin{minipage}{.48\linewidth}
        \centering
        \includegraphics[width=\linewidth]{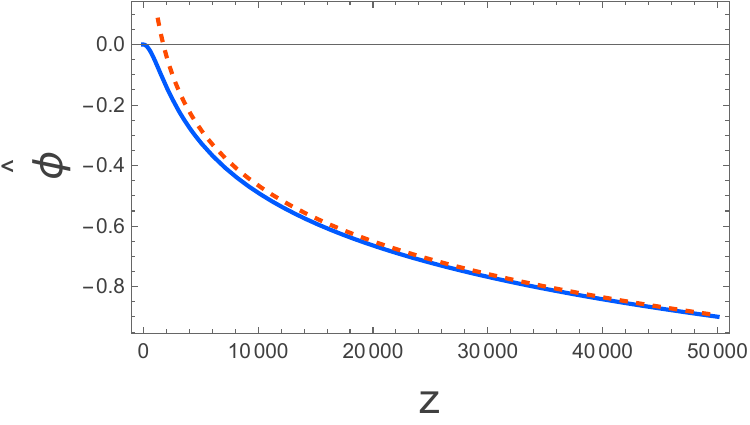}
    \end{minipage}
    \hfill
    \begin{minipage}{.48\linewidth}
        \centering
        \includegraphics[width=\linewidth]{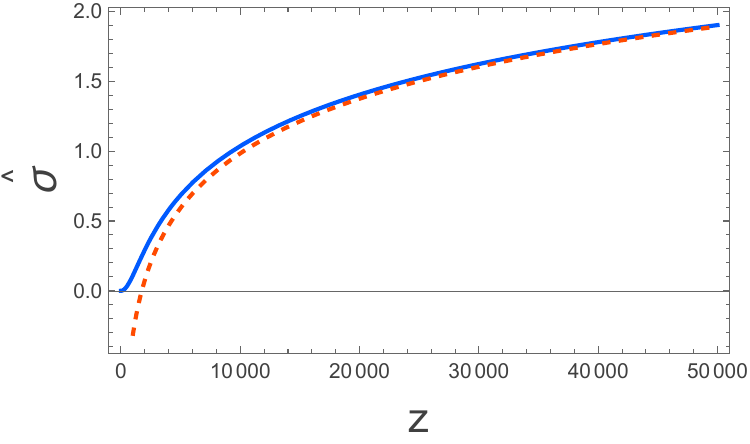}
    \end{minipage}
    \caption{Canonical solutions for $A$, $R$, $\phi$, and $\sigma$ (blue solid curves).
    The red dashed curves represent the solutions of the equations without the Casimir contribution, to which the geometry asymptotically approaches in the far region.}
    \label{fig:canonical_solutions}
\end{figure}

As the Casimir energy disappears at the far region, the solution asymptotically approaches the analytic solutions without Casimir energy.
The red dashed curves represent the asymptotic solutions \eqref{eq:sol_without_Casimir} with appropriately fitted parameters. In particular, the fitting parameters $\gamma$ and $\delta$, which determine the spacetime geometry in the asymptotic region, are important, and we obtain
\begin{align}
    \gamma = 0.89, \hspace{10pt} \delta = 2\pi - 0.44 %~[\text{rad}]
    . \label{eq:asymptotic_parameters}
\end{align}
General solutions are obtained from this solution following \eqref{eq:general_solution_Casimir}.
The transformations in \eqref{eq:general_solution_Casimir} change the normalization constants and the additive constants appearing in the far-region solution \eqref{eq:sol_without_Casimir}, but they do not change the exponents $\gamma$ and $\delta$. Therefore, the values of $\gamma$ and $\delta$ extracted from the canonical solution are universal within the regular finite-horizon branch.
We therefore conclude that, within our ansatz, Casimir backreaction does not cloak the classical naked 7-brane core in a way compatible with asymptotically locally flat asymptotics.

So far, we have considered the case in which the horizon is located at a finite value of the proper distance. Here, we comment on the case in which the horizon is pushed to infinite proper distance.
When the horizon is pushed to infinity in proper distance ($z = -\infty$), a natural candidate for the near-horizon region is an AdS-type metric. In this case, as $z \to -\infty$, one has
\begin{align}
    A(z) = e^{z/l_\text{AdS}}, \quad R(z) = R_h, 
    \quad \phi(z) = \phi_h, \quad \sigma(z) = \sigma_h.
\end{align}
$R_h,\ \phi_h,\ \sigma_h \in \mathbb{R}$. From the boundary condition \eqref{eq:boundary_condition_2} and horizon regularity condition \eqref{eq:boundary_condition_3}, the leading corrections to $R$, $\phi$, and $\sigma$ are of the same order as $A\sim e^{z/l_\text{AdS}}$. We then check whether these fields satisfy the equations of motion in the near-horizon limit. We focus on the fourth equation in \eqref{eq:eom_with_Casimir}. If $R_h \neq 0$, then we have
\begin{align}
    \dfrac{5\alpha}{2} \dfrac{e^{-2\sqrt{7}\sigma + 5\phi/2}}{R^{10}} \to \mathcal{O}(1), \quad \left(\dfrac{A^\prime}{A} + \dfrac{R^\prime}{R}\right)\phi^\prime + \phi^{\prime\prime} \to \mathcal{O}(e^{z/l_\text{AdS}}),
\end{align}
which is incompatible with the equation in the limit $z \to -\infty$. If $R_h = 0$, then we observe
\begin{align}
    \dfrac{5\alpha}{2} \dfrac{e^{-2\sqrt{7}\sigma + 5\phi/2}}{R^{10}} \to \mathcal{O}(e^{-5z/l_\text{AdS}}), \quad \left(\dfrac{A^\prime}{A} + \dfrac{R^\prime}{R}\right)\phi^\prime + \phi^{\prime\prime} \to \mathcal{O}(e^{z/l_\text{AdS}}).
\end{align}
Again, the equation is not satisfied in the limit $z \to -\infty$. Therefore, the solution does not admit an AdS-type near-horizon throat.

%%%%%%%%%%%%%%%%%%%%%%%%%%%%%%%%%%%%%%%%%%%%%%%%%%%%%%%%%%%%%%%%%%%%%%%%%%%%%%
\section{String theory applications}\label{sec:sols}
We now apply the general analysis of Secs.~\ref{sec:eom} and \ref{sec:solutions} to several string-theoretic 7-branes. In each case, the monodromy determines the boundary conditions of 10d fields and hence the coefficient $\alpha$ of the Casimir energy.

\subsection{R7-brane in IIB and IIA}

The bosonic sector of the type IIB supergravity action possesses an $SL(2,\mathbb{Z})$ duality symmetry.
In F-theory \cite{Vafa:1996xn}, this duality symmetry corresponds to modular transformations of the auxiliary torus that preserve its orientation.
The duality symmetry can be generalized to a transformation group that includes orientation-reversing transformations of the F-theory torus and also accounts for the fact that fermions transform under the double cover of the duality group. The resulting duality group is $GL^+(2, \mathbb{Z})$ \cite{Pantev:2016nze,Tachikawa:2018njr}.
When considering compactification on an $S^1$, one finds 
\begin{align}
    \Omega_1(GL^+(2, \mathbb{Z})) = \mathbb{Z}_2 \times \mathbb{Z}_2.
\end{align}
By the cobordism conjecture \cite{McNamara:2019rup}, this suggests the existence of branes charged under the corresponding nontrivial cobordism classes. In particular, one of the $\mathbb{Z}_2$ factors corresponds to orientation reversal of the F-theory torus, and the 7-brane carrying this charge is known as the reflection 7-brane (R7-brane) \cite{Dierigl:2022reg,Debray:2023yrs}.

Let us focus on the reflection generated by $(-1)^{F_L}$. In this case, the boundary conditions for 10d fields around the R7-brane are
\begin{align}
    \text{NSNS} : +1, \hspace{10pt} \text{RNS} : -1, \hspace{10pt} \text{NSR} : +1, \hspace{10pt} \text{RR} : -1
\end{align}
for the periodic spin structure on the $S^1$. However, in type IIB supergravity, the NSNS and RNS sectors, as well as the RR and NSR sectors, contain the same numbers of physical degrees of freedom (all sectors have 64 degrees of freedom). According to the expression for the Casimir energy \eqref{eq:Casimir_energy_for_sugra}, this implies
\begin{align}
    V_\text{Casimir} = 0.
\end{align}
This corresponds to the fact that the theory compactified on $S^1$ with the $(-1)^{F_L}$ twist (i.e., the Dabholkar--Park background \cite{Dabholkar:1996pc}) preserves half-maximal supersymmetry. 

Therefore, the R7-brane background is described, within our ansatz and away from the brane core, by the solution \eqref{eq:sol_without_Casimir}. In particular, the $\gamma=0$ branch gives an asymptotically locally flat region with an arbitrary deficit angle, and the origin is a naked conical singularity.

We have implicitly assumed that the axion field $C_0$ is trivial (i.e., constant) in the above solution.\footnote{See \cite{Cavusoglu:2026xiv} for analogous solutions in 4d with nontrivial axion profiles where axisymmetry is inevitably broken.}
For this assumption to be compatible with the R7-brane monodromy, the allowed values are
\begin{align}
    C_0 = 0,\,\dfrac{1}{2}.
\end{align}
Let $\theta$ denote the angular coordinate around the brane. The monodromy
maps $C_0$ to $-C_0$. The configuration $C_0=0$ is trivially compatible with
the monodromy. The constant profile $C_0=1/2$ is also compatible once the
axion shift symmetry $C_0\to C_0+1$ is taken into account, since
\begin{align}
  C_0(\theta=0)
  =
  - C_0(\theta=2\pi)
  \sim
  1-C_0(\theta=2\pi).
\end{align}
Note that the $C_0=0$ and $C_0=1/2$ 7-branes are viewed as boundaries of the Dabholkar--Park background~\cite{Dabholkar:1996pc} and the background identified in \cite{ParraDeFreitas:2022wnz,Montero:2022vva} (see also \cite{Bedroya:2021fbu}), respectively.

Although type IIA string theory does not possess an $SL(2,\mathbb Z)$-type duality symmetry, it does admit the discrete symmetry $(-1)^{F_L}$. Compactification on $S^1$ with $(-1)^{F_L}$ holonomy defines a nontrivial duality-twisted background, representing a nontrivial bordism class. The cobordism conjecture therefore suggests the existence of a 7-brane whose monodromy implements the $(-1)^{F_L}$ action.
The supergravity description of the IIA 7-brane is the same as in the type IIB case. The far-region behavior is therefore described by \eqref{eq:sol_without_Casimir}.

\subsection{Heterotic 7-brane}

In heterotic string theory, the existence of a 7-brane carrying an exotic charge was proposed in \cite{Kaidi:2023tqo,Kaidi:2024cbx}.\footnote{Similar codimension-2 branes in 9d and 8d heterotic strings were identified in \cite{Hamada:2024cdd,Hamada:2025cpe}.} We consider the heterotic string theory with gauge group $G = (E_8 \times E_8) \rtimes \mathbb{Z}_2$, where the $\mathbb{Z}_2$ factor corresponds to the exchange symmetry between the two $E_8$ groups. This exchange symmetry gives rise to a nontrivial disconnected sector of the gauge group, 
\begin{align}
    \pi_0(G) = \mathbb{Z}_2.
\end{align}
One can then consider a nontrivial gauge holonomy on $S^1$, which takes values in the disconnected component of $G$ and realizes a twist exchanging the two $E_8$ factors. In particular, within the $S^1$-compactified theory, background gauge-field configurations with trivial and nontrivial holonomy belong to distinct bordism classes. The cobordism conjecture therefore predicts the existence of a 7-brane whose monodromy implements this nontrivial holonomy. We now apply the general results above to this brane. 

The monodromy of the gauge field around the 7-brane is explicitly described as
\begin{align}
    A(x+2\pi R) = A^\prime (x), \hspace{10pt} A^\prime(x+2\pi R) = A (x).
\end{align}
Here, $A$ denotes the gauge field associated with one of the $E_8$ factors, while $A^\prime$ denotes the gauge field associated with the other $E_8$. We now reorganize the gauge-field degrees of freedom as follows:
\begin{align}
    A_+ \equiv A + A^\prime, \hspace{10pt} A_- \equiv A - A^\prime.
\end{align}
Then, the monodromies of these degrees of freedom are
\begin{align}
    A_+(x+2\pi R) &= A(x+2\pi R) + A^\prime(x+2\pi R) \nonumber\\
    &= A^\prime(x)+A(x) = A_+(x) : \text{periodic}, \\
    A_-(x+2\pi R) &= A(x+2\pi R) - A^\prime(x+2\pi R) \nonumber\\
    &= A^\prime(x)-A(x) = -A_-(x) : \text{antiperiodic}. 
\end{align}
Similarly, in the fermionic (gaugino) sector, half of the degrees of freedom obey periodic monodromy, while the other half obey antiperiodic monodromy. Since the Casimir contributions cancel whenever bosonic and fermionic degrees of freedom satisfy the same monodromy in \eqref{eq:Casimir_energy_for_sugra}, the Casimir contribution from the Yang-Mills gauge sector vanishes.
Therefore, only the degrees of freedom in the supergravity multiplet may contribute to the Casimir energy, whose numbers of bosonic and fermionic degrees of freedom are both 64. 
For the periodic spin structure on $S^1$, the 7-brane is viewed as a boundary of the CHL string theory \cite{Chaudhuri:1995fk}. The CHL string preserves all the supersymmetries of the original heterotic string theory, and the Casimir energy vanishes in this case. 
As in the R7-brane case, the solution is given by \eqref{eq:sol_without_Casimir}, which is expected to be valid away from the brane.
On the other hand, for the antiperiodic spin structure on $S^1$ (which corresponds to the situation in \cite{Kaidi:2023tqo,Kaidi:2024cbx}), the Casimir energy associated with this monodromy is given by
\begin{align}
    V_{\text{Casimir}}(e^{\frac{\sqrt{7}}{4}\sigma} R) = -\dfrac{3}{\pi^{15}} \left(\sum_{n \in \text{odd}} \dfrac{1}{n^{10}}\right) \dfrac{e^{-\frac{5\sqrt{7}}{2}\sigma}}{R^{10}}
    =-\dfrac{3}{\pi^{15}} \left(1 - \dfrac{1}{2^{10}}\right)\zeta(10) \dfrac{e^{-\frac{5\sqrt{7}}{2}\sigma}}{R^{10}}.
\end{align}
Thus the corresponding value of $\alpha$ is
\begin{align}
    \alpha_{\text{het}}
    =\dfrac{3}{\pi^{15}} \left(1 - \dfrac{1}{2^{10}}\right)\zeta(10),
\end{align}
which is nothing but the value of $\alpha$ chosen for the numerical solution shown in Fig.~\ref{fig:canonical_solutions}.
As we have seen there, all regular finite-horizon solutions in the present ansatz are not asymptotically locally flat.
This statement does not exclude more general completions involving nontrivial gauge-field profiles, non-axisymmetric configurations, or a would-be horizon located at infinite proper distance.

\section{Conclusions}

We have studied axisymmetric 7-brane backgrounds in the dilaton-gravity
sector of 10d supergravity, including the Casimir energy generated
by monodromies around the transverse circle. Within our ansatz, the problem
reduces to 3d Einstein gravity coupled to two scalar fields,
with or without a Casimir potential depending on the supersymmetry of the
associated circle compactification.

In the absence of Casimir energy, the equations are analytically solvable. The
resulting family of solutions admits asymptotically locally flat regions with
arbitrary deficit angles, and therefore provides the far-region geometry for
type II $(-1)^{F_L}$ reflection 7-branes and for the heterotic 7-brane with
periodic spin structure. However, these classical solutions have a naked singularity.

For nonzero Casimir energy, we asked whether this naked core can be replaced
by a regular finite-horizon black-brane geometry while preserving
asymptotically locally flat behavior. 
The radial equations have scaling properties that reduce the regular finite-horizon branch to a universal canonical solution. Our numerical construction shows that this canonical solution approaches, in the far region, a classical branch with fixed nonzero
scalar logarithmic running rather than the locally flat branch. Hence, within
the axisymmetric ansatz with trivial gauge-field backgrounds and constant
axion profiles, Casimir backreaction obstructs an
asymptotically locally flat finite-horizon black-brane completion of the naked 7-brane solution.
This obstruction is relevant for the heterotic 7-brane with antiperiodic spin
structure, for which the supergravity multiplet gives a nonzero Casimir
energy. By contrast, in the type II $(-1)^{F_L}$ reflection-brane case and in
the heterotic periodic-spin-structure case, the Casimir energy cancels.

There are several possible ways in which the conclusion could be modified beyond
the present setup. We have restricted ourselves to the axisymmetric dilaton-gravity
ansatz with trivial gauge-field backgrounds and constant axion profiles. Allowing
nontrivial gauge-field profiles, axion gradients, or non-axisymmetric configurations
may therefore change the conclusion.
We have also assumed that the regular horizon is located at finite
proper distance. At the end of Sec.~3.2, we have considered a simple infinite-distance
near-horizon ansatz of AdS type, and found that it is not compatible with the equations of motion. This observation, however, does not exclude more general infinite-distance completions, for example throats with running scalar fields. It would be interesting to classify such infinite-distance asymptotics and determine whether any of them can provide an asymptotically locally flat completion of the non-supersymmetric 7-branes considered here.

%%%%%%%%%%%%%%%%%%%%%%%%%%%%%%%%%%%%%%%%%%%%%%%%%%%%%%%%%%%%%%%%%%%%%%%%%%%%%%

%%%%%%%%%%%%%%%%%%%%%%%%%%%%%%%%%%%%%%%%%%%%%%%%%%%%%%%%%%%%%%%%%%%%%%%%%%%%%%
\subsection*{Acknowledgments}
We thank Yu Hamada and Hirotaka Yoshino for collaborations on a related project.
Y.H.\ was supported in part by JSPS KAKENHI Grant Nos.\ JP24H00976, JP24K07035, JP24KF0167, and by JST BOOST Program Japan Grant No.\ JPMJBY25E1.
H.K.\ was supported by JST SPRING, Japan Grant Number JPMJSP2104.

%%%%%%%%%%%%%%%%%%%%%%%%%%%%%%%%%%%%%%%%%%%%%%%%%%%%%%%%%%%%%%%%%%%%%%%%%%%%%%

\appendix

\section{Metric ansatz for general dimensional branes} \label{sec:Appendix_1}
In this section, following \cite{Horowitz:1991cd}, we generalize our ansatz without nontrivial gauge-field configurations to brane solutions in arbitrary dimensions and show that the system can be reduced to a lower-dimensional Einstein--Hilbert action coupled to scalar fields. We begin with the dilaton-gravity action.
\begin{align}
    S = \int d^{10} x \sqrt{-g}e^{-2\phi}[\mathcal{R}+4\partial_\mu\phi\partial^\mu\phi]. \label{eq:10dim_sugra_action_appendix}
\end{align}
First, we consider the following metric ansatz for general $p$-brane background solutions:
\begin{align}
    ds^2 = \hat{g}_{\mu\nu}dx^\mu dx^\nu + e^{2\psi}dx_idx^i, \quad \mu, \nu = 0,1, \ldots 9-p, \quad i = 9-p+1, \ldots, 9, \label{eq:string_frame_metric_ansatz_appendix_1}
\end{align}
where $x^i$ and $x^\mu$ denote the coordinates along and transverse to the brane, respectively. Within this ansatz, the relationship between the Ricci scalars of the ten-dimensional metric \eqref{eq:string_frame_metric_ansatz_appendix_1} and the $(10-p)$-dimensional metric $\hat{g}$ is
\begin{align}
    \mathcal{R} = \hat{\mathcal{R}} - ((\delta^i_i)^2 +\delta^i_i)\partial_\mu\psi\partial^\mu\psi  - 2\delta^i_i \hat{\nabla}^2 \psi. \color{black}
\end{align}
Furthermore, let us consider the Weyl transformation
\begin{align}
    \hat{g}_{\mu\nu} = e^{2\omega} \tilde{g}_{\mu\nu}.
\end{align}
Using the transformation formula for the Ricci scalar under a Weyl transformation,
\begin{align}
    \hat{\mathcal{R}} &= e^{-2\omega}\{\mathcal{\tilde{R}}-2(9-p)\tilde{\nabla}^2\omega -(8-p)(9-p)\tilde{\partial}_\mu\omega\tilde{\partial}^\mu\omega\},
\end{align}
one obtains
\begin{align}
    \mathcal{R} &= e^{-2\omega}\{\mathcal{\tilde{R}}-2(9-p)\tilde{\nabla}^2\omega -(8-p)(9-p)\tilde{\partial}_\mu\omega\tilde{\partial}^\mu\omega \nonumber \\ 
    & \hspace{40pt} - (p^2 + p)\tilde{\partial}_\mu\psi\tilde{\partial}^\mu\psi  - 2p \tilde{\nabla}^2 \psi - 2p(8-p)\partial_\mu \psi \partial^\mu \omega\}.
\end{align}
Thus, under the metric ansatz
\begin{align}
    ds^2 = e^{2\omega} \tilde{g}_{\mu\nu}dx^\mu dx^\nu + e^{2\psi}dx_idx^i, \hspace{10pt} \mu, \nu = 0,1, \ldots 9-p, \hspace{10pt} i = 9-p+1, \ldots, 9, \label{eq:string_frame_metric_ansatz_appendix_2}
\end{align}
the action \eqref{eq:10dim_sugra_action_appendix} becomes
\begin{align}
     S = &\int d^{10}x \sqrt{-\tilde{g}}e^{-2\phi -2\omega + p\psi +(10-p)\omega} [\mathcal{\tilde{R}}-2(9-p)\tilde{\nabla}^2\omega \nonumber \\
    & \hspace{30pt} -(8-p)(9-p)\tilde{\partial}_\mu\omega\tilde{\partial}^\mu\omega 
      - (p^2 + p)\tilde{\partial}_\mu\psi\tilde{\partial}^\mu\psi  \nonumber \\
      & \hspace{30pt} - 2p \tilde{\nabla}^2 \psi - 2p(8-p)\tilde{\partial}_\mu \psi \tilde{\partial}^\mu \omega + 4\tilde{\partial}_\mu\phi\tilde{\partial}^\mu\phi]. \label{eq:10dim_sugra_action_with_ansatz_general_p}
\end{align}
We now reparametrize the fields as follows for $p < 8$, in the same manner as in \cite{Horowitz:1991cd}.
\begin{align}
    \beta \phi &= \rho \dfrac{(p-3)}{2} - \sigma \dfrac{(7-p)}{2}\left[\dfrac{p}{8-p}\right]^{1/2}, \\
    \beta (2\omega) &= \rho \left[- \dfrac{6-p}{8-p}\right] - \sigma \left[\dfrac{p}{8-p}\right]^{1/2}, \\
    \beta (2\psi) &= \rho + \sigma \dfrac{p-6}{[p(8-p)]^{1/2}}, \\
    \beta &= -\left[ 2 \dfrac{9-p}{8-p}\right]^{1/2}. 
\end{align}
Then, the action \eqref{eq:10dim_sugra_action_with_ansatz_general_p} becomes
\begin{align}
   S = \int d^{10} x \sqrt{-\tilde{g}}\left[\tilde{\mathcal{R}}-\dfrac{1}{2}(\tilde{\nabla} \rho)^2 - \dfrac{1}{2}(\tilde{\nabla} \sigma)^2  \right]. 
\end{align}
In particular, the exponential factor in \eqref{eq:10dim_sugra_action_with_ansatz_general_p} disappears, and the $\tilde{\nabla}^{2}\ast$ term reduces to a surface term and can therefore be neglected. Moreover, if each field component is independent of the coordinates parallel to the brane, the integration over those directions can be carried out explicitly. For $p=7$, this reproduces our metric ansatz and parameters for the 7-brane.

%%%%%%%%%%%%%%%%%%%%%%%%%%%%%%%%%%%%%%%%%%%%%%%%%%%%%%%%%%%%%%%%%%%%%%%%%%%%%%

\section{Casimir energy in supergravity} \label{sec:Appendix_2}
In this appendix, we extract the Casimir energy from the effective potential including the one-loop effects of the field theory and compute it as a function of the radial coordinate in the presence of a black 7-brane at the center of spacetime. 

We begin with a field theory in $d$-dimensional spacetime described by a set of fields $\varphi$. To incorporate the one-loop effects in the partition function, it is sufficient to include the contribution from the quadratic fluctuation operator. Expanding the partition function with sources $J$ around a classical background configuration $\varphi_c$ of the fields, one obtains
\begin{align}
    Z[J] & \overset{\text{1-loop}}{=} e^{i(S[\varphi_c]+J \cdot \varphi_c)}\int \mathcal{D}\varphi_q e^{\frac{i}{2}\varphi_q \cdot \mathcal{O} \cdot \varphi_q} \nonumber \\
    & \hspace{8pt} = e^{i(S[\varphi_c]+J \cdot \varphi_c+\frac{i}{2}\sum_i \Tr \ln \mathcal{O}_i - i\sum_a \Tr \ln \mathcal{O}_a)}. \label{eq:partition_function}
\end{align}
Here, $i$ runs over the bosonic degrees of freedom, while $a$ runs over the fermionic degrees of freedom. $\mathcal{O}$ denotes the fluctuation operator,
\begin{align}
    \mathcal{O}_i \equiv -(\delta/\delta \varphi_i)(\delta S[\varphi]/\delta \varphi_i)|_{\varphi=\varphi_c}.
\end{align}
Equation \eqref{eq:partition_function} implies that the one-loop effective potential for the theory is
\begin{align}
    \Gamma[\varphi_c] = S[\varphi_c]+\frac{i}{2}\sum_{i\in \text{boson}} \Tr \ln \mathcal{O}_i - i\sum_{a \in \text{fermion}} \Tr \ln \mathcal{O}_a.
\end{align}
To compute the Casimir energy, it is sufficient to consider only the free-field part of the action in the fluctuation operator. Namely, for a scalar field degree of freedom with mass $m$,
\begin{align}
    \mathcal{O} \propto -\partial^2 + m^2,
\end{align}
and in that case, evaluating $\Tr \ln \mathcal{O}$ gives
\begin{align}
    \Tr \ln \mathcal{O} &= \int d^dx \bra{x} \ln (-\partial^2 + m^2)\ket{x} \nonumber \\
    &= \int d^dx \int d^dp d^dk \braket{x|p} \bra{p} \ln (-\partial^2+m^2) \ket{k} \braket{k|x} \nonumber \\
    &= \int d^dx \int \dfrac{d^d k}{(2\pi)^d} \ln (k^2 + m^2) \nonumber \\
    &= i\int d^dx \int \dfrac{d^d k}{(2\pi)^d} \ln (k_0^2 + k_1^2 + \ldots + k_{d-1}^2 + m^2).
\end{align}
In the final line, we have performed a Wick rotation. If the theory consists of a single real scalar field, one finds
\begin{align}
    \Gamma[\varphi] = &\int d^dx \left[(\text{kinetic term}) \right. \nonumber \\
    & - \left. \left\{V_0 + \dfrac{1}{2}\int \dfrac{d^dk}{(2\pi)^d}\ln (k_0^2 + k_1^2 + \ldots + k_{d-1}^2 + m^2)  + \dfrac{1}{2}\int \dfrac{d^dk}{(2\pi)^d}\ln \alpha \right\}\right].
\end{align}
Here, in the curly brackets, $V_0$ denotes the potential at the tree level, while the second term represents the Casimir potential. The final term arises from the coefficient multiplying $\partial^2$ in $\mathcal{O}$; however, if this coefficient is merely an $x$-independent constant, it does not contribute to the potential.

Now suppose that spacetime contains a defect surrounded by an $S^1$, so that the defect effectively creates a ``hole'' in spacetime. In this situation, the Casimir energy should be evaluated on each $S^1$ of radius $R$ centered around the defect. In particular, one of the momentum directions becomes discretized into Kaluza--Klein modes:
\begin{align}
    V_\text{Casimir} = \dfrac{1}{2} \sum_{n = -\infty}^{\infty} \dfrac{1}{2\pi R}\int \dfrac{d^{d-1}k}{(2\pi)^{d-1}}\ln \left(k_0^2 + k_1^2 + \ldots + k_{d-2}^2 + m^2 + \dfrac{(n+\theta)^2}{R^2}\right).
\end{align}
Here, $\theta$ determines the boundary condition of the field around the $S^1$. This integral can be evaluated as follows \cite{Hamada:2017yji}:
\begin{align}
    V_\text{Casimir} &= \dfrac{1}{2} \sum_{n = -\infty}^{\infty} \dfrac{1}{2\pi R}\int \dfrac{d^{d-1}k}{(2\pi)^{d-1}}\ln \left(k_0^2 + k_1^2 + \ldots + k_{d-2}^2 + m^2 + \dfrac{(n+\theta)^2}{R^2}\right) \nonumber \\
    &= -\dfrac{1}{2} \dfrac{d}{ds}\sum_{n = -\infty}^{\infty} \dfrac{1}{2\pi R}\int \dfrac{d^{d-1}k}{(2\pi)^{d-1}} \left. \left(k_0^2 + k_1^2 + \ldots + k_{d-2}^2 + m^2 + \dfrac{(n+\theta)^2}{R^2}\right)^{-s} \right|_{s=0} \nonumber \\
    &= -\dfrac{1}{2} \sum_{n = -\infty}^{\infty} \dfrac{1}{2\pi R} (4\pi)^{-(d-1)/2} \dfrac{d}{ds} \left[\dfrac{\Gamma\left(s - \frac{d-1}{2}\right)}{\Gamma(s)} \left(\dfrac{(n+\theta)^2}{R^2} + m^2\right)^{-(s-(d-1)/2)} \right]_{s=0}
\end{align}
In the final line, we used the standard dimensional-regularization formula, as reviewed for example in \cite{Peskin:1995ev}. For small $s$,
\begin{align}
    \Gamma(s) \sim \dfrac{1}{s} - \gamma + \mathcal{O}(s).
\end{align}
and therefore
\begin{align}
    \dfrac{\Gamma\left(s - \frac{d-1}{2}\right)}{\Gamma(s)} \sim \dfrac{\Gamma(s-\frac{d-1}{2})}{1/s - \gamma} \sim \Gamma\left(s-\frac{d-1}{2}\right) (1+\gamma s) s.
\end{align}
Hence, 
\begin{align}
    \left.\dfrac{\Gamma\left(s - \frac{d-1}{2}\right)}{\Gamma(s)} \right|_{s=0} = 0, \quad \dfrac{d}{ds} \left(\dfrac{\Gamma\left(s - \frac{d-1}{2}\right)}{\Gamma(s)}\right)_{s=0} = \Gamma\left(- \frac{d-1}{2}\right).
\end{align}
Using this, we obtain
\begin{align}
    V_\text{Casimir} &= -\dfrac{1}{2}  \dfrac{1}{2\pi R} (4\pi)^{-(d-1)/2} \Gamma\left(- \frac{d-1}{2}\right) \sum_{n = -\infty}^{\infty}\left(\dfrac{(n+\theta)^2}{R^2} + m^2\right)^{(d-1)/2} \nonumber \\
    &= -\dfrac{1}{2} \dfrac{1}{2\pi R^d} (4\pi)^{-(d-1)/2} \Gamma\left(- \frac{d-1}{2}\right) F\left(-\frac{d-1}{2}; \theta, mR \right).
\end{align}
The function $F$ can further be evaluated as \cite{Ponton:2001hq}
\begin{align}
    F\left(-\frac{d-1}{2}; \theta, mR \right) = &\dfrac{\sqrt{\pi}}{\Gamma(-\frac{d-1}{2})}(mR)^{d} \left(\Gamma\left(-\dfrac{d}{2}\right) \right. \nonumber \\
    &\left. \hspace{30pt} + 4\sum_{p=1}^{\infty} (\pi p mR)^{-\frac{d}{2}} \cos(2\pi p \theta) K_{-\frac{d}{2}}(2\pi p mR) \right).
\end{align}
Substituting this expression yields
\begin{align}
    V_\text{Casimir} &= -\dfrac{1}{2}(4\pi)^{-\frac{d}{2}} m^d \Gamma\left(-\dfrac{d}{2}\right) -2 \dfrac{1}{(2\pi)^d} \left(\dfrac{m}{R}\right)^{\frac{d}{2}} \sum_{n=1}^\infty \dfrac{\cos(2\pi n \theta)}{n^{\frac{d}{2}}} K_{\frac{d}{2}} (2\pi n mR). \label{eq:d_dim_Casimir_diverge_appendix}
\end{align}
In particular, the divergent first term is precisely equal to the Casimir energy $V_\text{Casimir}^{(0)}$ of a scalar field in spacetime without defect: 
\begin{align}
    V_\text{Casimir}^{(0)} &= \dfrac{1}{2} \int \dfrac{d^d k}{(2\pi)^d} \ln (k_0^2 + k_1^2 + \ldots + k_{d-1}^2 + m^2) \nonumber \\
    &= -\dfrac{1}{2}\dfrac{d}{ds} \left. \int \dfrac{d^d k}{(2\pi)^d} (k_0^2 + k_1^2 + \ldots + k_{d-1}^2 + m^2)^{-s} \right|_{s=0} \nonumber \\
    &= -\dfrac{1}{2} (4\pi)^{-\frac{d}{2}} \dfrac{d}{ds} \left (\dfrac{\Gamma(s-\frac{d}{2})}{\Gamma(s)}m^{d-2s} \right)_{s=0} \nonumber \\
    &= -\dfrac{1}{2} (4\pi)^{-\frac{d}{2}} m^{d} \Gamma\left(-\frac{d}{2}\right).
\end{align}
This term is canceled by a (divergent) cosmological-constant counterterm. Therefore, only the second term in \eqref{eq:d_dim_Casimir_diverge_appendix} is physically meaningful:
\begin{align}
    V_{\text{Casimir}} = -2 \dfrac{1}{(2\pi)^d} \left(\dfrac{m}{R}\right)^{\frac{d}{2}} \sum_{n=1}^\infty \dfrac{\cos(2\pi n \theta)}{n^{\frac{d}{2}}} K_{\frac{d}{2}} (2\pi n mR).
\end{align}
For $d=4$, this result is consistent with \cite{Hamada:2017yji}.

For fermions, one similarly finds
\begin{align}
    \mathcal{O}^2 = -\partial^2 + m^2
\end{align}
Therefore, combining all contributions, the full Casimir energy is given in terms of the number of physical degrees of freedom $n_p$, the spin $s_p$, and the mass $m_p$ of each particle species $p$ as follows:
\begin{align}
    V_{\text{Casimir}} = \sum_{p} (-1)^{2s_p+1} \dfrac{2 n_p}{(2\pi)^{d}} \left(\dfrac{m_p}{R}\right)^{\frac{d}{2}} \sum_{n=1}^\infty \dfrac{\cos(2\pi n \theta_p)}{n^{\frac{d}{2}}} K_{\frac{d}{2}} (2\pi n m_pR).
\end{align}
We are particularly interested in the case where $d=10$ and all field degrees of freedom are massless. Using the asymptotic form of the Bessel function 
\begin{align}
    K_n(x) \sim 2^{n-1}(n-1)! \, x^{-n} \quad (x\to0),
\end{align}
our Casimir energy becomes
\begin{align}
     V_\text{Casimir} = \sum_{p} (-1)^{2s_p+1} \dfrac{768 n_p}{(2\pi)^{15}}\dfrac{1}{R^{10}} \sum_{n=1}^\infty \dfrac{\cos(2\pi n \theta_p)}{n^{10}}. 
\end{align}
For example, the supergravity multiplet does not contribute when the bosonic and fermionic degrees of freedom obey the same monodromy.
For an antiperiodic spin structure, one instead obtains
\begin{align}
    V_\text{Casimir}^{(\text{sugra})} &= - \dfrac{768 \times 64}{(2\pi)^{15}}\dfrac{1}{R^{10}} \sum_{n=1}^\infty \dfrac{1-\cos(\pi n)}{n^{10}} \nonumber \\
    &= - \dfrac{768 \times 64}{(2\pi)^{15}}\dfrac{1}{R^{10}} \sum_{n\in \text{odd}}^\infty \dfrac{2}{n^{10}} \nonumber \\
    &=- \dfrac{3}{\pi^{15}} \left(\sum_{n\in \text{odd}}^\infty \dfrac{1}{n^{10}}\right)\dfrac{1}{R^{10}}. 
\end{align}
%%%%%%%%%%%%%%%%%%%%%%%%%%%%%%%%%%%%%%%%%%%%%%%%%%%%%%%%%%%%%%%%%%%%%%%%%%%%%%

\section{Analytical solutions for equations of motion without Casimir energy} \label{sec:Appendix_3}

In this section, under our metric ansatz, we analytically solve the equations of motion without the Casimir contribution given in \eqref{eq:eom_without_Casimir}. 
Using the second equation in \eqref{eq:eom_without_Casimir}, one can eliminate $\sigma$ and $\phi$ from the first equation in \eqref{eq:eom_without_Casimir} and the third equation in \eqref{eq:eom_without_Casimir}, yielding
\begin{align}
    A'R' + AR'' &= (AR')' = 0, \\
    A'R' + A''R &= (A'R)' = 0.
\end{align}
Integrating each of them, we obtain
\begin{align}
    AR' = c_1, \quad A'R = c_2, \quad c_1, c_2 \in \mathbb{R}. \label{eq:integrated_eom}
\end{align}
Adding them together and performing a further integration, we obtain
\begin{align}
    AR' + A'R = (AR)' = c_1 + c_2 \quad \Rightarrow \quad AR = (c_1 + c_2)z + c_3, \quad c_3 \in \mathbb{R}. \label{eq:integrated_AR}
\end{align}
Here, for $(c_1 + c_2) \neq 0$, since the system of equations \eqref{eq:eom_without_Casimir} possesses a shift symmetry in $z$, one may use this symmetry to set $c_3 = 0$. Substituting this back into the second equation in \eqref{eq:integrated_eom}, one obtains
\begin{align}
    \dfrac{A'}{A} = \dfrac{c_2}{(c_1 + c_2)z},
\end{align}
which is solved as
\begin{align}
    |A| = e^{c_4}|(c_1 + c_2)z|^{\frac{c_2}{c_1 + c_2}}, \quad c_4 \in \mathbb{R}.
\end{align}
Combining this with \eqref{eq:integrated_AR}, we finally obtain, for $z>0$,
\begin{align}
    A = a_0 z^\gamma, \quad R = r_0 z^{1-\gamma},
\end{align}
where $\gamma = c_2 / (c_1 + c_2)$. Then, from the second equation in \eqref{eq:eom_without_Casimir}, 
\begin{align}
    \phi^{\prime 2} + \sigma^{\prime 2} = \dfrac{4\gamma(1-\gamma)}{z^2}. \label{eq:constraint_for_scalars}
\end{align}
Since the left-hand side is manifestly non-negative, $\gamma$ should satisfy $0 \leq \gamma \leq 1$. On the other hand, the fourth and fifth equations in \eqref{eq:eom_without_Casimir}, namely the equations of motion for the scalar fields, are
\begin{align}
    \dfrac{\phi^\prime}{z} + \phi^{\prime\prime} = 0, 
    \quad \dfrac{\sigma^\prime}{z} + \sigma^{\prime\prime} = 0 \quad \Rightarrow \quad \phi^\prime = \dfrac{k}{z}, \quad \sigma^\prime = \dfrac{l}{z}, \quad k,l \in \mathbb{R}. \label{eq:eom_for_scalars}
\end{align}
In order to satisfy the constraint \eqref{eq:constraint_for_scalars}, 
\begin{align}
    k^2 + l^2 = 4\gamma(1-\gamma),
\end{align}
and so the solutions for \eqref{eq:eom_for_scalars} are
\begin{align}
    \phi &= \phi_0 + 2\sqrt{\gamma(1-\gamma)}\sin\delta\ln z, \\
    \sigma &= \sigma_0 + 2\sqrt{\gamma(1-\gamma)}\cos\delta\ln z.
\end{align}
$\phi_0, \sigma_0 \in \mathbb{R}$ and $0 \leq \delta < 2\pi$. In summary, we obtain the following results.
\begin{align}
    g_{\mu\nu} &= e^{\frac{\phi}{2}+\frac{\sqrt{7}}{2}\sigma}[-a_0^2 z^{2\gamma}dt^2 + dz^2 + r_0^2 z^{2(1-\gamma)}d\varphi^2] + e^{\frac{\phi}{2} - \frac{\sigma}{2\sqrt{7}}}dx_i dx^i, \\
    \phi &= \phi_0 + 2\sqrt{\gamma(1-\gamma)}\sin\delta\ln z, \\
    \sigma &= \sigma_0 + 2\sqrt{\gamma(1-\gamma)}\cos\delta\ln z,
\end{align}
where $a_0, r_0 >0, ~0 \leq \gamma \leq 1, ~0 \leq \delta < 2\pi$. The shift symmetry in $z$ restores the replacement $z \to z-z_0$ used in Sec.~\ref{sec:solutions}, \eqref{eq:sol_without_Casimir}.

If $c_1 + c_2 = 0$ in \eqref{eq:integrated_AR}, $A'/A = \text{const.}$ and
\begin{align}
    A \propto e^{az}, \quad R \propto e^{-az}. 
\end{align}
However, substituting this into \eqref{eq:integrated_AR}, one obtains
\begin{align}
    \phi^{\prime 2} + \sigma^{\prime 2} = -4a^2,
\end{align}
and the non-negativity of the left-hand side implies that only $a=0$ is allowed. This corresponds to the trivial flat solution in which $A, R, \phi$, and $\sigma$ are all constant (in that case, $\varphi$ is not an angular coordinate). 

%%%%%%%%%%%%%%%%%%%%%%%%%%%%%%%%%%%%%%%%%%%%%%%%%%%%%%%%%%%%%%%%%%%%%%%%%%%%%%

\pagebreak
\bibliographystyle{JHEP}
\bibliography{reference} 
\end{document}